\documentclass{PoS}

\usepackage{amsmath}
\usepackage{microtype}

\title{Impact of CMS dijets in 5.02 TeV pPb and pp collisions on EPPS16 nuclear PDFs}

\ShortTitle{Impact of CMS dijets in 5.02 TeV pPb and pp collisions on EPPS16 nPDFs}

\author{Kari J. Eskola\\
       University of Jyvaskyla, Department of Physics, P.O. Box 35, FI-40014 University of Jyvaskyla, Finland \\
       Helsinki Institute of Physics, P.O. Box 64, FI-00014 University of Helsinki, Finland \\
       E-mail: \email{kari.eskola@jyu.fi}}

\author{\speaker{Petja Paakkinen}
        \\ University of Jyvaskyla, Department of Physics, P.O. Box 35, FI-40014 University of Jyvaskyla, Finland \\
        E-mail: \email{petja.paakkinen@jyu.fi}}

\author{Hannu Paukkunen\\
       University of Jyvaskyla, Department of Physics, P.O. Box 35, FI-40014 University of Jyvaskyla, Finland \\
       Helsinki Institute of Physics, P.O. Box 64, FI-00014 University of Helsinki, Finland \\
       E-mail: \email{hannu.paukkunen@jyu.fi}}

\abstract{The CMS measurement of dijet pseudorapidity distributions in pPb versus pp collisions at 5.02 TeV provides a direct probe on nuclear gluon PDFs. We show that while the predicted pPb pseudorapidity distributions suffer from sizable free-proton PDF uncertainties, the ratios of the pPb and pp distributions are practically insensitive to scale and free-proton PDF choices. We find the CMS data on pPb to pp ratios to be in good agreement with the EPPS16 nuclear modifications. Using a non-quadratic extension of the Hessian PDF reweighting method, we study the impact of these data on the EPPS16 nuclear PDFs. Relative to EPPS16, we find stronger evidence for mid-$x$ gluon antishadowing as well as indication for larger gluon shadowing at small $x$. The data are also able to further constrain the gluon PDF in the EMC region.}

\FullConference{International Conference on Hard and Electromagnetic Probes of High-Energy Nuclear Collisions\\
		30 September - 5 October 2018\\
		Aix-Les-Bains, Savoie, France}

\begin{document}

\section{Introduction}

In the most recent global analysis of nuclear parton distribution functions (nPDFs), EPPS16~\cite{Eskola:2016oht}, a wealth of new observables were introduced, enabling for a flavour-separated nPDF parameterization. Even with this most extensive set of data to date, direct constraints for nuclear gluon PDFs remained particularly scarce: the available data on PHENIX deuteron--gold $\pi^0$ production~\cite{Adler:2006wg} and CMS proton--lead (pPb) dijet forward-to-backward ratio~\cite{Chatrchyan:2014hqa} were able to constrain the nuclear gluon modifications mainly in the antishadowing region, leaving everything below $x \sim 10^{-2}$ practically unconstrained at low $Q^2$. Moreover, these measurements with gold and lead ions give very little hint over the mass-number dependence.

Thanks to the reference 5.02 TeV proton--proton (pp) data taking, the CMS collaboration has been able to perform the first measurement of a dijet pPb to pp ratio~\cite{Sirunyan:2018qel}, with sensitivity to gluon PDF nuclear modifications in a lead nucleus down to $x \sim 3 \cdot 10^{-3}$. Here, we demonstrate the impact of these data on the EPPS16 nPDFs, after sorting out an issue with the compatibility of these data with the free-proton PDFs used in the EPPS16 analysis. For the nPDF part, we employ an improved version of the Hessian PDF reweighting~\cite{Paukkunen:2014zia}, taking into account first non-quadratic elements in the $\chi^2$-function of the original fit. A more detailed discussion on the method and its application on this particular case will be presented in a future publication~\cite{dijets}.

\section{Issue with proton PDFs}

In Figure~\ref{fig:pp_and_pPb}, the normalized dijet spectra in pp and pPb collisions as measured by CMS are shown and compared with NLO theory predictions obtained with the NLOJet++ code~\cite{Nagy:2003tz}. For these calculations, the CT14 NLO PDFs~\cite{Dulat:2015mca} were used, multiplied with the central set of EPPS16 nuclear modification factors for the incoming lead ion. Here one observes a problem: the predicted pp spectra are wider than the measured distributions, the data points lying for the most part even outside the CT14 PDF uncertainties. A similar deviation is observed with the pPb spectra, hinting that the problem has a common source.

To test whether the issue could be solved with suitable modifications in the proton PDFs, we have performed a Hessian PDF reweighting on the CT14 PDFs with the pp data. The resulting modifications of CT14 gluons are shown in Figure~\ref{fig:CT14} for two values of global tolerance: $\Delta\chi^2 = 100$, which is the nominal tolerance used in the CT14 analysis, and an artificially low $\Delta\chi^2 = 10$, which would in a global fit correspond to giving additional weight to these data. For both cases, the obtained modifications are significant, especially at large $x$, where a large suppression with respect to the original central set would be preferred. Similar results have recently been found in reweighting CT14 with top-production data~\cite{Azizi:2018iiq}, and hence seeing such effects here is not entirely unexpected.

\begin{floatingfigure}[t]
  \includegraphics[width=\textwidth]{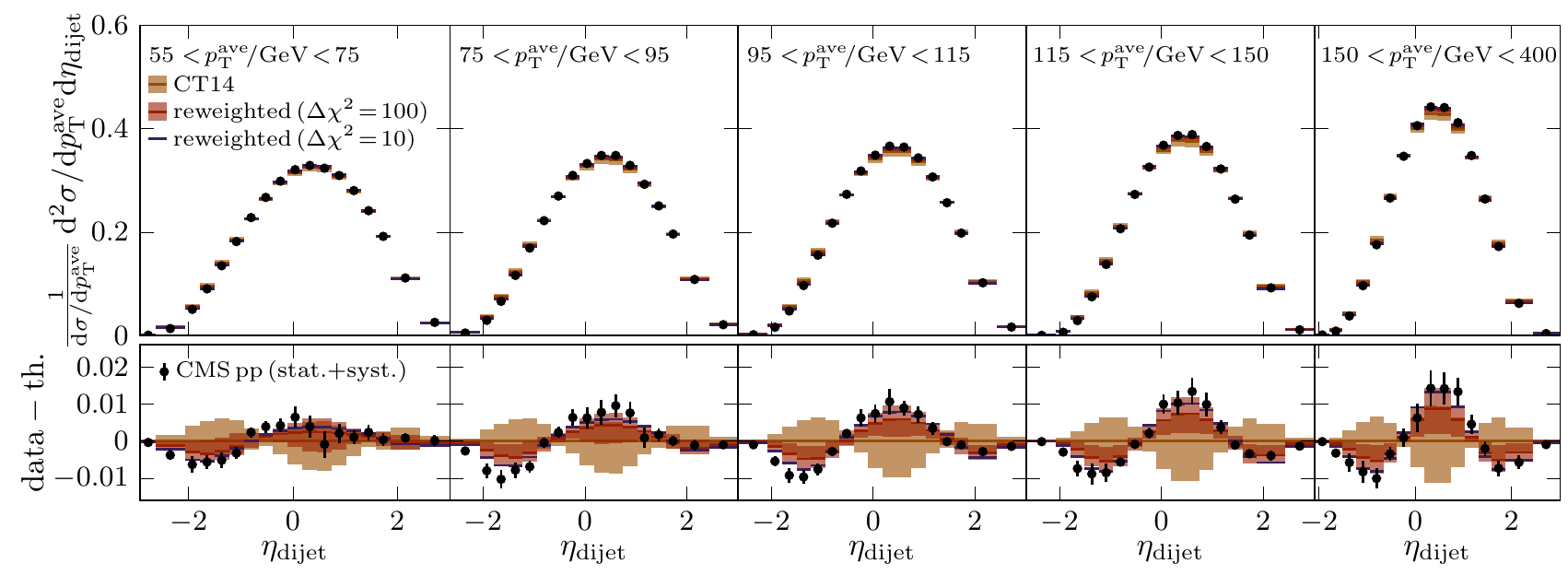}
  \includegraphics[width=\textwidth]{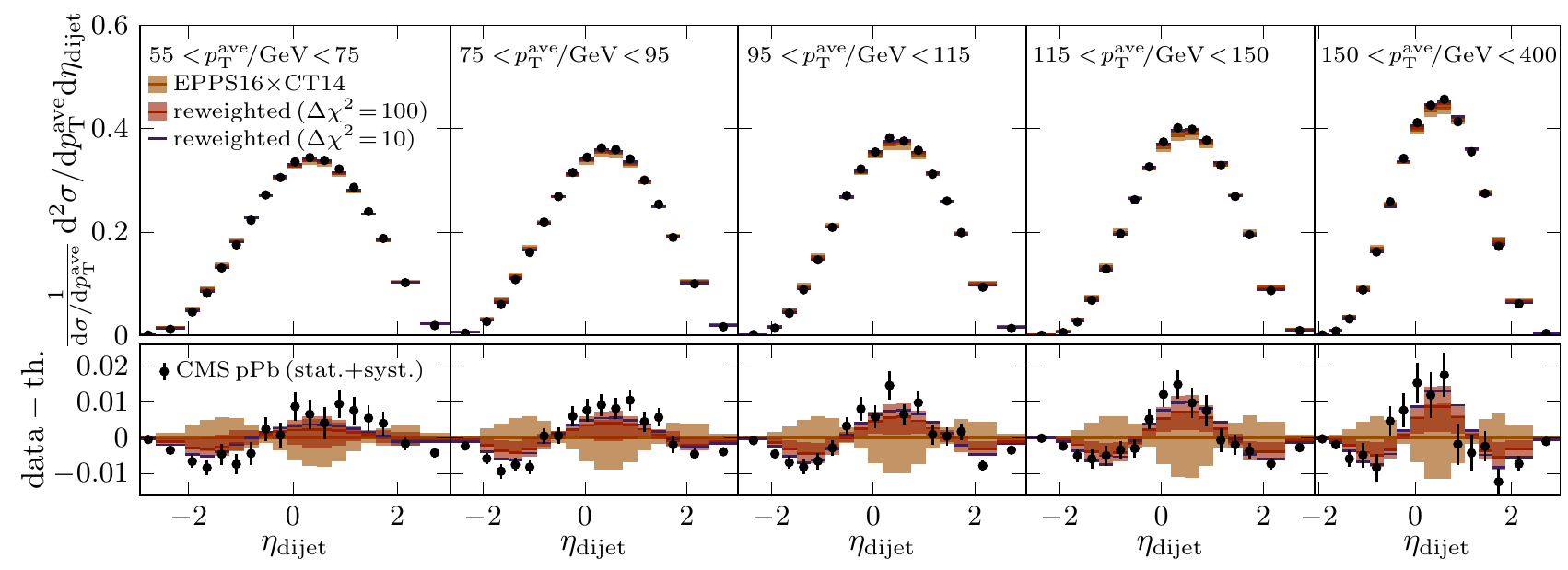}
  \vspace{-0.6cm}
  \caption{The CMS dijet rapidity spectra in pp (\emph{top}) and pPb (\emph{bottom}) collisions compared with NLO predictions using the CT14 PDFs. Results for CT14 PDFs reweighted with the pp spectra are also shown for two different choices of global tolerance. In the pPb predictions the central EPPS16 nuclear modifications were used.}
  \label{fig:pp_and_pPb}
\end{floatingfigure}

In Figure~\ref{fig:pp_and_pPb}, also the impact of the reweighting on the predictions is presented. The reweighting clearly improves the agreement with the pp data. Still, at $\eta_{\rm dijet} \sim -2$, the reweighting is not able to fully reproduce the data, even when applied with the additional weight ($\Delta\chi^2 = 10$). This might indicate a parametrization issue in the CT14, although NNLO effects could also become important in this region. As the experimental uncertainties here are systematics dominated, it could also happen that a shift in some of the experimental parameters would improve the agreement. We will study this possibility once the data correlations become publicly available.

\begin{floatingfigure}
  \begin{minipage}{0.39\textwidth}
    \centering
    \includegraphics[width=1.05\textwidth]{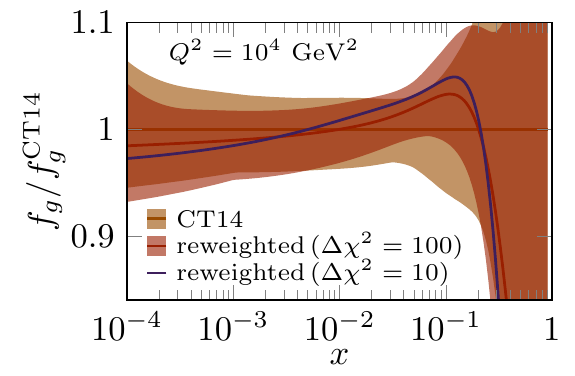}
  \end{minipage}
  \caption{The modification of CT14 gluon PDF under reweighting with the pp data.}
  \label{fig:CT14}
\end{floatingfigure}

As in the case of pp spectra, also the predictions for pPb spectra get a strong impact from the reweighting. This observation is of great importance: had we used these data directly in a nuclear PDF analysis with the original CT14 proton PDFs, we would have ended up overestimating the nuclear effects. What differs from the pp case is that here we have additional deviation at $\eta_{\rm dijet} \sim 2$, with the predictions overshooting the data. We will next come to understand this as an indication of stronger shadowing for gluons than what we have in the central set of EPPS16 PDFs.

\section{Nuclear modification ratio and impact on EPPS16}

Taking the ratio of the normalized pPb and pp spectra discussed above, the proton PDF uncertainties effectively cancel. This is evident from Figure~\ref{fig:RpPb}, where the uncertainties from the CT14 PDFs, as shown in the top panels, are clearly negligibly small. After this cancellation of proton-PDF effects, the data and EPPS16 are in good accordance. We also note that the factor-two scale variations are small, indicating that higher order corrections would not have large impact on this observable, and leaving the nuclear modifications as the dominant source of theoretical uncertainty.

\begin{floatingfigure}[t]
  \includegraphics[width=\textwidth]{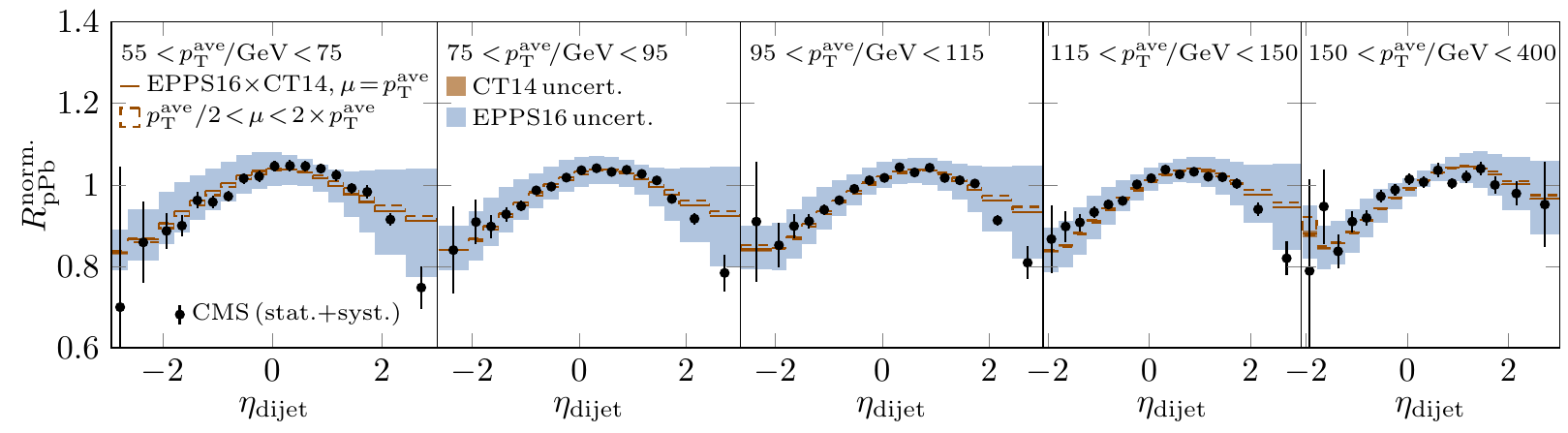}
  \includegraphics[width=\textwidth]{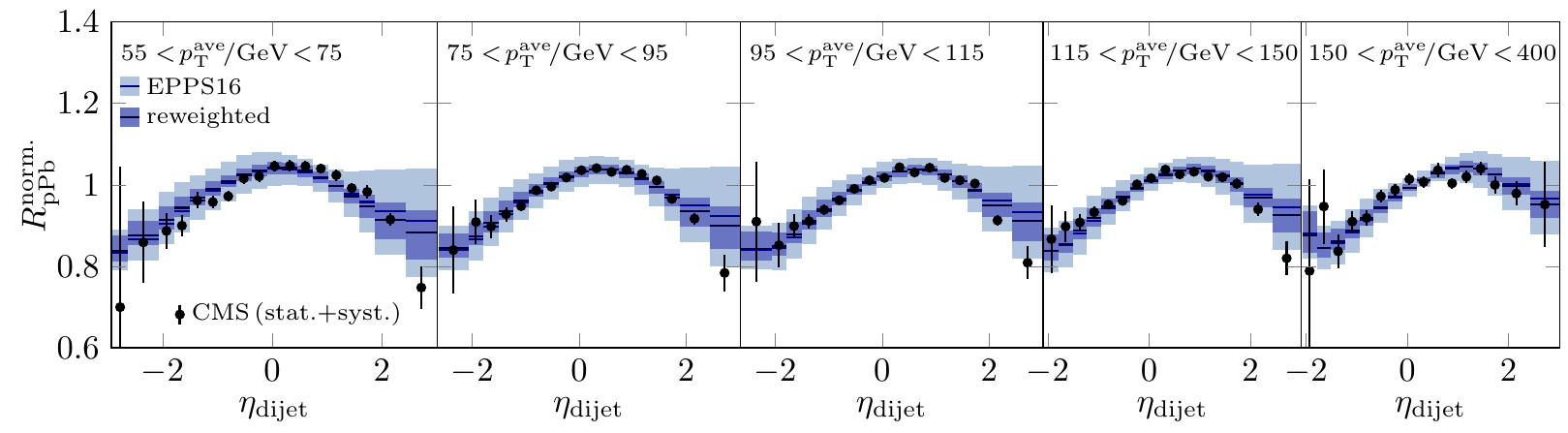}
  \vspace{-0.6cm}
  \caption{The CMS normalized dijet nuclear modification ratio compared with NLO predictions using the EPPS16 nuclear modifications. The CT14 and scale uncertainties (\emph{top}) and the reweighted results (\emph{bottom}) are shown.}
  \label{fig:RpPb}
\end{floatingfigure}

\begin{floatingfigure}
  \begin{minipage}{0.39\textwidth}
    \centering
    \includegraphics[width=1.05\textwidth]{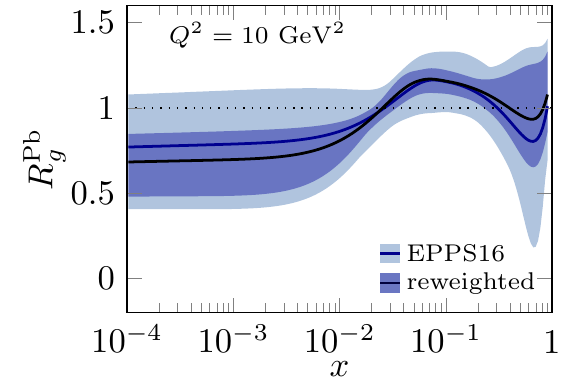}
  \end{minipage}
  \caption{The effect of reweighting with the nuclear modification ratio on EPPS16 gluons for the lead nucleus.}
  \label{fig:EPPS16}
\end{floatingfigure}

With these observations, it appears well justified to use the dijet $R_{\rm pPb}^{\rm norm.}$ to further constrain the EPPS16 nPDFs. The effect on predictions are shown in the bottom panels of Figure~\ref{fig:RpPb}, where a vast reduction in the EPPS16 uncertainties can be seen. The reweighted results nicely match with the data, except for the systematically outlying data points at the most forward rapidity bins. Comparing with Figure~\ref{fig:EPPS16}, where the impact on EPPS16 gluon modifications are shown, we observe that matching with these data points would require extremely deep nuclear shadowing. Such a strong shadowing might even contradict the measurements with forward D-meson production~\cite{Aaij:2017gcy}, which seem to suggest a rather similar shadowing pattern as observed here~\cite{Kusina:2017gkz,Dmeson}.

That aside, we observe that the data put stringent constraints on the gluon nuclear modifications, more than halving the antishadowing uncertainty. As anticipated above, a downward pull at small $x$ is evident, giving strong support for gluon shadowing. Also the EMC region gets better constrained, but the evidence for high-$x$ suppression remains inconclusive.

\section{Conclusions}

We have shown here that to obtain good agreement with CMS pp dijet measurements, the CT14 proton PDFs need to be significantly modified. The modifications needed in CT14 to describe pp data have also a large impact on pPb predictions. This once again emphasizes the fact that it is imperative to understand the pp baseline before making far-reaching conclusions from the pPb data. Upon taking the ratio of the pPb and pp spectra, one becomes effectively sensitive only to the nuclear modifications of the PDFs. We have shown that the data on the dijet nuclear modification ratio is able to significantly constrain the gluon nuclear modifications, giving strong evidence for mid-$x$ antishadowing and for a rather strong small-$x$ shadowing.

\acknowledgments

The authors have received funding from the Academy of Finland, Project 297058 of K.J.E.\ and 308301 of H.P.;
P.P.\ acknowledges the financial support from the Magnus Ehrnrooth Foundation.
The cross-section calculations were performed within a computing cluster of the Finnish IT Center for Science (CSC) under the Project jyy2580.

\end{document}